\documentclass[conference]{IEEEtran}
\IEEEoverridecommandlockouts
\usepackage{cite}
\usepackage{amsmath,amssymb,amsfonts}
\usepackage{bm}
\usepackage{algorithmic}
\usepackage{graphicx}
\usepackage{textcomp}
\usepackage{xcolor}
\usepackage{enumerate}
\usepackage{siunitx}
\usepackage{balance}
\usepackage{caption}
\usepackage{subcaption}
\def\BibTeX{{\rm B\kern-.05em{\sc i\kern-.025em b}\kern-.08em
    T\kern-.1667em\lower.7ex\hbox{E}\kern-.125emX}}

\allowdisplaybreaks

\newcommand{\comment}[1]{}

\newcommand{\textsub}[2]{#1_{ \mathrm{#2}}}
\newcommand{\textsup}[2]{#1^{ \hspace{1 pt} \mathrm{#2}}}

\newcommand{\pp}{\bm{p}} 
\newcommand{\RR}{\bm{R}} 

\newcommand{\Rtil}{\tilde{\bm{R}}}
\newcommand{\arraypos}{\bm{\Delta}} 

\newcommand{\zerovec}{\bm{0}}
\newcommand{\yy}{\bm{y}}
\newcommand{\YY}{\bm{Y}}
\renewcommand{\aa}{\bm{a}}
\newcommand{\bb}{\bm{b}}
\newcommand{\taz}{\textsup{\theta}{az}}
\newcommand{\tel}{\textsup{\theta}{el}}
\newcommand{\dpseu}{\breve{d}}
\newcommand{\xx}{\bm{x}}
\newcommand{\ZZ}{\bm{Z}}
\newcommand{\zz}{\bm{z}}
\newcommand{\kk}{\mathbf{k}}

\newcommand{\param}{\bm{\eta}}

\renewcommand{\ss}{\bm{s}}
\newcommand{\alal}{\bm{\alpha}}
\newcommand{\II}{\bm{I}}
\newcommand{\xixi}{\bm{\xi}}

\newcommand{\vecobsmean}[1]{\bm{\mu}_{#1}}
\newcommand{\JJ}{\bm{J}}

\newcommand{\CC}{\bm{C}}
\newcommand{\cc}{\bm{c}}
\newcommand{\FF}{\bm{F}}
\newcommand{\CF}{\bm{\Omega}}

\newcommand{\gamgam}{\bm{\gamma}}

\newcommand{\transpose}[1]{#1^{T}}

\newcommand{\myexp}[1]{e^{\hspace{1 pt} \displaystyle{#1}}}
\newcommand{\expect}[1]{\mathbb{E} \hspace{-1.5 pt} \left[ #1 \right]}

\newcommand{\theReals}{\mathbb{R}}
\newcommand{\theComplex}{\mathbb{C}}
\newcommand{\identity}[1]{\mathbb{I}_{#1}}

\begin{document}

\bstctlcite{IEEEexample:BSTcontrol}

\title{Optimized Vehicular Antenna Placement for Phase-Coherent Positioning
\\
\thanks{This work was partially supported by the Wallenberg AI, Autonomous Systems and Software
Program (WASP) funded by the Knut and Alice Wallenberg Foundation, and 
the Swedish Research Council (Grants 2022-03007 and 2024-04390).}
}

\author{Victor Pettersson\IEEEauthorrefmark{1,2}, Musa Furkan Keskin\IEEEauthorrefmark{1}, Carina Marcus\IEEEauthorrefmark{2}, Henk Wymeersch\IEEEauthorrefmark{1} 
\\
\IEEEauthorrefmark{1}Chalmers University of Technology, Gothenburg, Sweden,
\IEEEauthorrefmark{2}Magna Electronics, Vårgårda, Sweden\\
email: victor.pettersson@magna.com}

\maketitle

\begin{abstract}

Distributed multi-antenna systems are an important enabling technology for future intelligent transportation systems (ITS), showing promising performance in vehicular communications and near-field (NF) localization applications. This work investigates optimal deployments of phase-coherent sub-arrays on a vehicle for NF localization in terms of a Cramér-Rao lower bound (CRLB)-based metric. Sub-array placements consider practical geometrical constraints on a three-dimensional vehicle model accounting for self-occlusions. Results show that, for coherent NF localization of the vehicle, the aperture spanned by the sub-arrays should be maximized and a larger number of sub-arrays results in more even coverage over the vehicle orientations under a fixed total number of antenna elements, contrasting with the outcomes of incoherent localization. Moreover, while coherent NF processing significantly enhances accuracy, it also leads to more intricate cost functions, necessitating computationally more complex algorithms than incoherent processing. 

\end{abstract}


\section{Introduction}

The vision for future intelligent transportation systems (ITS) includes sharing of safety-critical information, like sensor- and positioning data, among road users and infrastructure. The aim is to extend the reach of individual vehicle's perception and to improve the overall safety of traffic participants \cite{ETSI_102_638, V2X_Pos_Opportunities, V2X_Network_Design, V2X_Sidelink_01, ISAC_1}. Thus, some ITS functions put strict requirements on positioning accuracy which motivates exploring alternatives to contemporary solutions such as stand-alone global navigation satellite systems (GNSS) \cite{V2X_Network_Design}. 

For ITS, an enabling technology is the vehicle's antenna system. It is expected that conventional roof-top antenna deployments will not meet the requirements of ITS and car manufacturers are exploring alternative multi-antenna distributed deployments \cite{V2XChannels_AntennaPlacement, V2XChannels_AntennaPlacement_2}.
In \cite{V2XChannels_AntennaPlacement_2}, it is shown that distributed antenna deployments on the vehicle body enhance communication reliability through spatial diversity and mitigates shadowing effects. Distributed antenna deployments can also aid in vehicular positioning: in \cite{V2X_Sidelink_01}, a distributed deployment of antenna elements on the vehicle front bumper enables precise relative positioning of ahead vehicles in a simulated vehicle-to-vehicle (V2V) platooning scenario. 

From a deployment optimization standpoint, vehicle-side antenna deployments have previously focused on enhancing \textit{communication} capabilities \cite{V2XChannels_AntennaPlacement_2}. 
The \textit{positioning} performance of vehicles equipped with large or distributed arrays has been explored \cite{V2X_Sidelink_01}, particularly under near-field conditions that account for wavefront curvature, though without optimizing the deployment.
In fact, deployment optimization for positioning has been mainly considered in the context of  \textit{infrastructure} deployments \cite{DeploymentOpt_1, DeploymentOpt_2, V2X_Network_Design}. 
Existing research on \textit{user-side} deployments generally targets \textit{far-field} conditions and two-way propagation, such as in vehicular radar systems; for instance, \cite{sparseArrayDesign_TSP_2019} optimizes sparse sub-array placements to achieve desired beam-pattern characteristics. Another study, \cite{mimo_opt_geo_TSP_2024}, optimizes sparse planar arrays for vehicular radar in the far-field, focusing on minimizing angular ambiguity and enhancing resolution. 
Since distributed arrays on a vehicle can be phase-locked by distributing a reference oscillator signal via cables, \textit{near-field} effects should be considered, both for positioning and communication purposes. 

\begin{figure}
    \centering
    \includegraphics[width=0.8\linewidth]{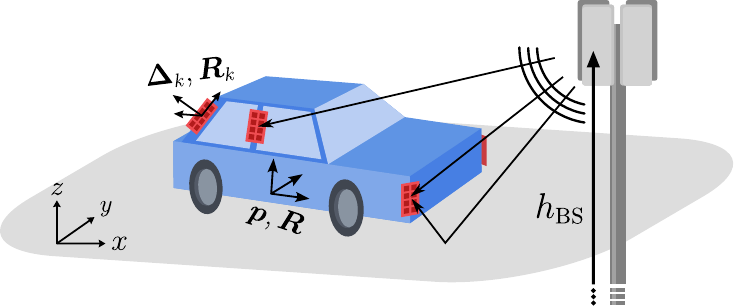}
    \caption{Illustration of the considered downlink localization scenario.}
    \label{fig:SystemModelIllustration}
\end{figure}

This paper seeks to bridge this research gap by exploring \textit{user-side}, vehicular deployment of \textit{phase-coherent}, distributed antennas for a \textit{near-field} positioning system. Our contributions are as follows: 
\begin{enumerate} 
\item We investigate near-field positioning from the user-side perspective, introducing a novel vehicular positioning setup. 
\item We present optimal antenna deployments for a vehicular 5G millimeter-wave (FR2) downlink scenario that includes considerations for vehicle self-occlusion effects and specular multipath. 
\item We establish fundamental guidelines for the localization-optimal deployment of an array of sub-arrays on a vehicle, accommodating both phase-coherent and phase-incoherent operations. 
\end{enumerate}

\section{System Model \& Problem Formulation} \label{sec:SystemModel}

A vehicle located  in proximity of a base-station (BS) constitutes the considered scenario illustrated in Fig.~\ref{fig:SystemModelIllustration}. 
\subsection{Geometry Model}
The BS is modeled as a single emitting source in the origin of a global coordinate system. The ground is a flat plane characterized by the point $\transpose{[0, 0, -h_{\mathrm{BS}}]}$, where $h_\mathrm{BS}$ is the BS height above ground, and surface normal $\transpose{[0, 0, 1]}$. Let $\pp = \transpose{[x, y, z]} \in \theReals ^ {3 \times 1}$ denote the vehicle position where $z$ is known. Known is also the vehicle orientation, represented by an SO(3) rotation matrix $\RR \in \theReals ^ {3 \times 3}$ that maps the vehicle-local coordinate system to the global coordinate system.

The vehicle is equipped with $K$ planar antenna arrays referred to as ``sub-arrays" \cite{sparseArrayDesign_TSP_2019} numbered $k = 1, \dots, K$, each with known placement $\arraypos_k$ expressed in the vehicle-local coordinate system and $\arraypos_k$ is restricted to lie on the surface of the vehicle body. The global position of the $k$-th array is $\pp_k = \pp + \RR \arraypos_k$. The sub-arrays posses their own rotations defined by $K$ known SO(3) rotation matrices $\RR_k$, mapping the local coordinate system of the $k$-th array to the vehicle-local coordinate system. Let $\tilde{\bm{u}}_k$ be a point in space expressed by a set of coordinates defined in the $k$-th array-local system, then the same point is \begin{equation} \label{eq:GlobalLocalCoordRelation}
    \bm{u} = \pp_k + \Rtil_k \tilde{\bm{u}}_k
\end{equation} in the global system where $\Rtil_k = \RR \RR_k$. 
Considering the $k$-th sub-array, apart from its position and orientation, it is also characterized by its element positions $\mathcal{Q}_k = \{\tilde{\bm{q}}_{k, m}\}_{m=1}^{M_k}$ expressed in the coordinate system of the $k$-th sub-array. 

\subsection{Signal Model} \label{sec:SignalModel}

The BS employs an orthogonal frequency division multiplexing (OFDM) modulation scheme and transmits a number of known pilot symbols $\xx \in \theComplex ^{N \times 1}$ over $N$ subcarriers, e.g., as in the case of 5G sounding reference signals \cite{uplinkPosRS_ICC_2023, 6D_Loc_TVT_2023}. Observations are made at each (fully digital) sub-array over $M_k$ elements. The signal incident on the $k$-th sub-array travels through a number of paths indexed by $\ell$, where $\ell = 0$ specifically indicates the LOS path and $\ell \geq 1$ denotes any NLOS paths. NLOS paths are assumed single-bounce and reflections are specular in nature. The set of path indexes is $\mathcal{L}_k \subseteq \mathbb{N}$ and its elements may vary between sub-arrays due to orientation-dependent vehicle self-occlusions.\footnote{By considering the scenario in Fig.~\ref{fig:SystemModelIllustration} and assuming no reflective objects in the vehicle's vicinity, it is understood that only the LOS path and a ground-reflected (GR) path are possibly incident on the sub-arrays. Let $\ell = 1$ indicate the GR path, this gives $\mathcal{L}_k \in \{\{\}, \{0\}, \{1\}, \{0, 1\} \}$ and refers to the fully occluded case, the LOS-only case, the GR-only case or the LOS+GR case, respectively.} The spatial-frequency domain observation matrix $\YY_{k} \in \theComplex^{M_k \times N}$ is \cite{uplinkPosRS_ICC_2023, NF_Sensing_CRB_2024, 6D_Loc_TVT_2023} \begin{equation} \label{eq:observation}
    \YY_{k} = \sideset{}{_{\ell \in \mathcal{L}_k}} \sum \alpha_{k, \ell} \myexp{ j \phi_{k, \ell}} \sqrt{\textsub{P}{tx}} \aa_{k, \ell} \transpose{(\bb_{k, \ell} \odot \xx)} + \ZZ_k \,,
\end{equation} 
where $\alpha_{k, \ell} \in \theReals$ and $\phi_{k, \ell} \in \theReals$ are the channel magnitude- and phase response for the $\ell$-th path, respectively, $\textsub{P}{tx}$ is the BS average transmit power, $\aa_{k, \ell} \in \theComplex ^{M_k \times 1}$ is the spatial steering vector, $\bb_{k, \ell} \in \theComplex ^{N \times 1}$ is the frequency-domain steering vector and $\ZZ_k \in \theComplex ^ {M_k \times N}$ is i.i.d. additive white Gaussian noise, meaning $\mathrm{vec}\left( \ZZ_k \right) \sim \mathcal{CN} \hspace{-1.5 pt} \left( \zerovec, \sigma^2 \identity{M\hspace{-1.5 pt} N} \right)$.

In \eqref{eq:observation}, it is further assumed that the duration of one OFDM symbol (given by $T_\mathrm{sym} = 1/\Delta_f + T_\mathrm{cp}$ where $T_{\mathrm{cp}}$ is the duration of a cyclic prefix) is sufficiently short so that time-variations of the channel can be ignored \cite{OFDMProcessing}.

\subsection{Geometrical Relations} \label{sec:SignalModel_GeomQuantities}

At individual sub-arrays we apply a local plane-wave far-field model. The direction of an approaching wave is described by the azimuth- and elevation angle-of-arrival defined by \begin{align}
    &\taz_{k, \ell} = -\mathrm{atan2} \left( [\tilde{\bm{u}}_{k, \ell}]_2, [\tilde{\bm{u}}_{k, \ell}]_1 \right) \label{eq:ThetaAz} \\
    &\tel_{k, \ell} = \mathrm{asin} \left( [\tilde{\bm{u}}_{k, \ell}]_3 / \| \tilde{\bm{u}}_{k, \ell}\| \right) \,,\label{eq:ThetaEl}
\end{align} respectively, for some source- or virtual source point $\tilde{\bm{u}}_{k, \ell}$ expressed in the local coordinates of the $k$-th sub-array. The source points follow \begin{equation} \label{eq:SourcePoints}
    \tilde{\bm{u}}_{k, \ell} = \begin{cases}
    -\transpose{\Rtil}_k \pp_k, & \hspace{3 pt} \text{for } \ell = 0  \\
    -\transpose{\Rtil}_k \left(\pp_k - \pp^v_{\ell} \right), & \hspace{3 pt} \text{for } \ell \geq 1
    \end{cases}
\end{equation} using \eqref{eq:GlobalLocalCoordRelation}, where $\pp^v_\ell$ is a virtual source point in the global coordinate system from mirroring the true source in the plane of the $\ell$-th reflecting surface \cite{YusPaper}. 

The spatial steering vector describes the phase progression over the sub-array's elements. With the above model and definitions, the spatial steering vector's elements are \begin{equation}
    \left[ \aa_{k, \ell} \right]_m = \left[\aa_k (\taz_{k, \ell}, \tel_{k, \ell}) \right]_{\hspace{-1pt}m} = \exp \hspace{-2pt} \left ( j \transpose{\tilde{\bm{q}}}_{k, m} \kk(\taz_{k, \ell}, \tel_{k, \ell}) \right ) \,,
\end{equation} for $m = 1, \dots, M_k$, where 
\begin{equation} \label{eq:wavenumber} \begin{split}
    & \kk(\taz_{k, \ell}, \tel_{k, \ell}) = \\
    & = \frac{-2\pi}{\lambda} [\cos\taz_{k, \ell} \cos \tel_{k, \ell}, \hspace{1pt} -\sin\taz_{k, \ell} \cos \tel_{k, \ell}, \hspace{1pt} \sin \tel_{k, \ell}]^T
\end{split}
\end{equation} 
is the wavenumber vector pointing in the direction of the incident wave.

The received signal experiences a delay equal to $\tau_{k, \ell} = \dpseu_{k \ell} / c$ where $c$ is the speed of light in vacuum and $\dpseu_{k, \ell}$ is the ``pseudo distance". The latter quantity is defined as \begin{equation} \label{eq:PseudoDist}
    \dpseu_{k, \ell} = d_{k, \ell} + \delta_d \,,
\end{equation} where $d_{k, \ell} = \| \tilde{\bm{u}}_{k, \ell} \|$ is the true propagation distance and $\delta_d$ is an offset due to an unknown timing offset $\delta_\tau = \delta_d / c$ between the vehicle and BS. With this, the frequency-domain steering vector's elements are \begin{equation} \label{eq:SteeringVecFreq}
    [\bb_{k, \ell}]_n = [\bb(\dpseu_{k, \ell})]_n = \exp ( -j 2\pi \Delta_f n \dpseu_{k, \ell} / c )
\end{equation} for subcarrier indices $n = 0, 1, \dots, N - 1$ and subcarrier spacing $\Delta_f$.

The channel phase responses $\phi_{k, \ell}$ are modeled according to \begin{equation}
        \phi_{k, \ell} = \begin{cases}
            -2\pi f_c d_{k, 0} / c + \delta_{\phi, k}, &\hspace{3 pt} \text{for } \ell = 0 \\ 
            -2\pi f_c d_{k, \ell} / c + \delta_{\phi, k} + \angle \Gamma_{k, \ell}, & \hspace{3 pt} \text{for } \ell \geq 1
        \end{cases} \,, \label{eq:PhaseResponse}
\end{equation} where $f_c$ is the transmitted waveform carrier frequency and $\delta_{\phi, k}$ is an unknown phase offset between the BS and the $k$-th sub-array. The parameter $\angle \Gamma_{k, \ell}$ represents an unknown phase shift from a single reflection with reflection coefficient $\Gamma_{k, \ell}$.

\subsubsection*{Two Modes of Operation}

We consider two modes of operation: (i) the first mode uses a common local oscillator for all sub-arrays and the system is phase-synchronized prior to performing the localization task. In this mode, there only exists a single offset between the vehicle and the BS, meaning $\delta_{\phi, k} = \delta_{\phi}$. Only then can phase information be exploited and the system is considered NF \cite{uplinkPosRS_ICC_2023}. The second mode, (ii), is without phase synchronization between sub-arrays. Hereafter, these modes are referred to as the \textit{coherent} and \textit{incoherent} modes, respectively.

\subsection{Problem Formulation}

Given $\{\YY_{k}\}_{k=1}^K$ in \eqref{eq:observation}, our goal is to determine the placement and orientation of sub-arrays on the vehicle to optimize the system localization performance (i.e., accuracy of estimation of $\pp$). In order to quantify the performance of a specific deployment, we consider
a metric for the general localization error, denoted $\rho$, which will be  formulated in Section \ref{sec:LocalizationMetric}. Let a specific deployment be characterized by the set $\mathcal{A} = \{\bm{\Delta}_k\}_{k=1}^{K} \cup \{\bm{R}_k\}_{k=1}^{K} \cup \{\mathcal{Q}_k\}_{k=1}^{K}$. A constraint on the number of sub-arrays $K$ is enforced, and the optimization problem with respect to $\mathcal{A}$ is formulated as \begin{equation} \label{eq:OrigOptProblem}
    \begin{split}
        \text{minimize}_{\mathcal{A}} \hspace{6 pt} & \rho (\mathcal{A}) \\
        \text{s.t.} \hspace{6 pt} & |\mathcal{A}| = K \,,
    \end{split} 
\end{equation} emphasizing that $\rho = \rho(\mathcal{A})$ and $|\cdot|$ denotes set cardinality.

\section{Proposed Metric and Method} \label{sec:LocalizationMetric}

\subsection{Proposed Metric}

For the purpose of tractability of the metric and usefulness in both NF and far-field, we propose to use a percentile of the CRLB (considering random vehicle locations and orientations). The scenario involves several distributed sub-arrays with individual orientations and visibilities under both coherent and incoherent operation. Hence, we first detail the CRLB derivation and then introduce the metric. 

\subsubsection{Cramér-Rao Lower Bound} \label{sec:CRLB}

From the received signal modeled by \eqref{eq:observation} at each sub-array, the system is tasked with jointly synchronizing and localizing itself relative to the BS by estimating the position $\transpose{[x, y]}$ as well as synchronization parameters in  \begin{equation} \ss = \begin{cases} \transpose{[\delta_d, \delta_{\phi}]} \in \theReals^{2 \times 1}, & \hspace{-2 pt} \text{coherent} \\\transpose{[ \delta_d, \delta_{\phi, 1}, \dots, \delta_{\phi, K}
]} \in \theReals^{(1 + K) \times 1}, & \hspace{-2 pt} \text{incoherent} \end{cases}
\end{equation} for the respective modes. Here, the index notation $k = 1, \dots, K$ is overloaded to specify only non-occluded sub-arrays, and is true for the rest of this section. The nuisance parameters are $\alal = \transpose{[\transpose{\alal}_\mathrm{LOS}, \transpose{\alal}_\mathrm{GR}]}$ where \begin{align}
    \alal_{\mathrm{LOS}} & = \transpose{[\alpha_{1, 0}, \dots , \alpha_{K_\mathrm{LOS}, 0}]} \\
    \alal_{\mathrm{GR}} & = \transpose{[\alpha_{1, 1}, \dots , \alpha_{K_\mathrm{GR}, 1}, \angle \Gamma_{1, 1}, \dots,  \angle \Gamma_{K_\mathrm{GR}, 1}]}
\end{align} and are necessarily indexed by separate path-specific schemes $k_\mathrm{LOS} = 1, \dots, K_\mathrm{LOS}$ and $k_\mathrm{GR} = 1, \dots, K_\mathrm{GR}$ for the LOS- and GR path, respectively. 

The unknown parameter vector is \begin{equation} 
\param = \transpose{[x, y, \transpose{\ss}, \transpose{\alal}
    ]} \in \theReals^{P \times 1}
\end{equation} with $P = 4 + K_\mathrm{LOS} + 2K_\mathrm{GR}$ or $3 + K + K_\mathrm{LOS} + 2K_\mathrm{GR}$ in the coherent and incoherent modes, respectively. The accuracy of an (unbiased) estimator $\hat{\param}$ of $\param$ is assessed by means of the CRLB. If $\bm{C}_{\hat{\param}} \in \theReals^{P \times P}$ denotes its covariance, then the CRLB of $\bm{C}_{\hat{\param}}$ is expressed \begin{equation}
    \bm{C}_{\hat{\param}} - \II_{\param}^{-1} \succeq \bm{0}
\end{equation} where ``$\succeq \bm{0}$" is applied in the positive semi-definite sense and $\II_{\param}$ is the Fisher information matrix (FIM) of $\param$ \cite{Kay}. 

Equation \eqref{eq:observation} is now rewritten using the $\mathrm{vec}(\cdot)$ operator: \begin{equation} \label{eq:VectorizedObs}
    \yy_k = \mathrm{vec}(\YY_k) = \sideset{}{_{\ell \in \mathcal{L}_k}} \sum \vecobsmean{k, \ell} + \zz_k \in \theComplex^{M_k \hspace{-1 pt} N \times 1}
\end{equation} where $\sum_\ell \vecobsmean{k, \ell} = \expect{\yy_k}$, \begin{equation} \label{eq:VectorizedObsSignalPart}
    \vecobsmean{k, \ell} = \alpha_{k, \ell} \myexp{j\phi_{k, \ell}} \sqrt{P_{\mathrm{tx}}} \hspace{-2 pt} \big( \bb(\dpseu_{k, \ell}) \odot \xx \big) \hspace{-2 pt} \otimes \hspace{-1 pt} \aa(\taz_{k, \ell}, \tel_{k, \ell}) \,,
\end{equation} and $\zz_k = \mathrm{vec}(\ZZ_k)$. Note that some of the parameters in $\param$ do not enter \eqref{eq:VectorizedObsSignalPart} and \eqref{eq:VectorizedObs} directly, but rather through another set of sub-array-specific channel parameters \begin{equation}
    \xixi_{k, \ell} =\transpose{[\taz_{k, \ell}, \tel_{k, \ell}, \dpseu_{k, \ell},  \phi_{k, \ell}, \alpha_{k, \ell}]} \in \theReals^{5 \times 1}.
\end{equation} The channel parameters are parameterized by $\param$ through \eqref{eq:ThetaAz}, \eqref{eq:ThetaEl}, \eqref{eq:PseudoDist} and \eqref{eq:PhaseResponse} through \eqref{eq:SourcePoints} as described in Section \ref{sec:SignalModel_GeomQuantities}. The total channel parameter vector $\xixi_k$ is the concatenation of $\xixi_{k, \ell}$ for $\ell \in \mathcal{L}_k$ and the corresponding FIM has elements \cite{Kay}
\begin{equation}
    \left[ \II_{\xixi_k} \right]_{i, j} = \frac{2}{\sigma^2}\Re \left \{ \frac{\partial \hspace{0.5 pt} \expect{\yy_k^H}}{\partial [\xixi_k]_{i}} \frac{\partial \hspace{0.5 pt} \expect{\yy_k}}{\partial [\xixi_k]_{j}} \right \}.
\end{equation} 

The total information conveyed from $K$ observations modeled by \eqref{eq:VectorizedObs} at each sub-array is \cite{Kay} \begin{equation}\label{eq:FIM}
\II_{\param} = \sum^K_{k=1} \transpose{\JJ}_k \II_{\xixi_k} \JJ_k \,,
\end{equation} considering that observations are independent also across sub-arrays where $\JJ_k = \partial \xixi_k / \partial \param$ and $\transpose{\JJ}_k \II_{\xixi_k} \JJ_k$ is the contribution of the $k$-th sub-array to $\II_{\param}$ \cite{TheoryOfPointEstimatesBook}. With this, the lower bound on the position-error magnitude, the position error bound (PEB), can be formulated as \begin{equation} \label{eq:PEB}
    \mathrm{PEB} = \sqrt{ \mathrm{tr} ( [\II^{-1}_{\param}]_{1:2, 1:2})} \hspace{1 pt}.
\end{equation}

\subsubsection{Proposed Metric $\rho$}

The localization error is, in addition to the sub-array placement $\mathcal{A}$, a function of the vehicle's orientation and the position itself. This work restricts the orientations of the vehicle to be rotations about the $z$-axis only, i.e, $\RR = \RR_z(\varphi)$ where $\varphi$ is the rotation angle referenced to the positive $x$-axis. For the single-BS geometry, it is noted that $\mathrm{PEB} = \mathrm{PEB}(\varphi, r; \mathcal{A})$, where $r = \sqrt{x^2 + y^2}$ is the radial distance from the BS along the ground. The metric $\rho$ for a specific deployment $\mathcal{A}$ is defined as \begin{equation} \label{eq:Rho}
    \rho \hspace{-1 pt} : \hspace{3 pt} \mathrm{Pr}\big( \mathrm{PEB}(\varphi, r; \mathcal{A}) < \rho \big) = 1-\varepsilon,
\end{equation} i.e., $\rho$ is the $(1-\varepsilon)$-th percentile value of the PEB when regarding $\varphi$ and $r$ as random variables. The value of $\varepsilon$ can be set based on the application requirements. 

\subsection{Proposed Optimization Methodology} \label{sec:OptMethod}

The optimization problem formulated in \eqref{eq:OrigOptProblem} based on the metric $\rho(\mathcal{A})$ defined in \eqref{eq:Rho} is not tractable and additional constraints are required. 

\subsubsection{Practical Sub-Array Deployments}
First, the set of possible deployments $\mathcal{A}$ is discretized by only allowing sub-arrays to be placed at finite number of pre-defined points on the vehicle body, referred to as grid points. Specifically, 
\begin{enumerate}[i)]
    \item Grid points specify the placement and orientation of sub-arrays. Placements are restricted to those deemed practical from a vehicle-manufacturer standpoint.

    \item Sub-array placements are symmetric and mirrored in a vertical plane intersecting the vehicle body along its center length-wise. If a grid point is placed center and in that plane, then it is not mirrored. Likewise, orientations are mirrored in the said plane.

    \item All sub-arrays in a grid are identical with respect to the element distribution such that $\mathcal{Q}_k = \mathcal{Q}$ and $M_k = M$. 
\end{enumerate} Two grids are used in this work with identical placements and sub-array rotations, shown in Fig.~\ref{fig:Grid}, but varying element distributions $\mathcal{Q}$. They vary between $2 \times 2$ ($M = 4$) and $4 \times 2$ ($M = 8$) elements and the latter configuration is the one in Fig.~\ref{fig:Grid}. There are 20 grid points in total, corresponding to 34 unique sub-array placements. The vehicle model is approximately 4.0 m $\times$ 1.8 m $\times$ 1.7 m.

{

\setlength{\abovecaptionskip}{3 pt}

\begin{figure}
\centering
 \begin{subfigure}[t]{0.44\columnwidth}
    \centering
    \includegraphics[width= \textwidth]{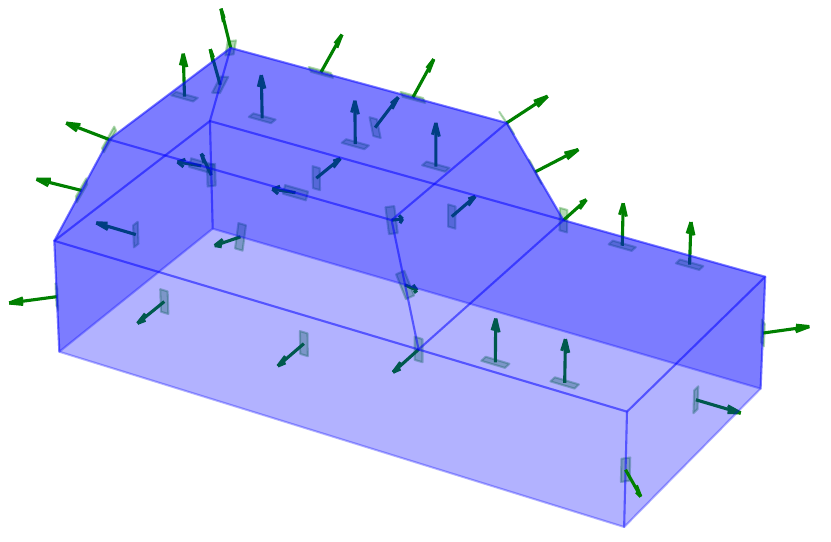}
    \caption{}
    \label{fig:Grid}
 \end{subfigure}
 \hfill
 \begin{subfigure}[t]{0.50\columnwidth}
    \centering
    \includegraphics[width= \textwidth]{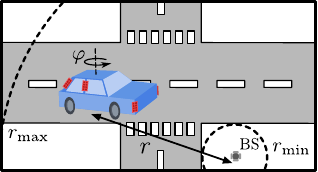}
    \caption{}
    \label{fig:RotIllustration}
 \end{subfigure}
 \caption{(a): Grid used for optimization trials. Sub-array orientations are indicated by arrows, sizes scaled up by a factor of ten. (b): Example traffic scenario for restricting $(\varphi, r)$ when sampling $\mathrm{PEB}(\varphi, r; \mathcal{A})$.}
\label{fig:GridAndRot}
\end{figure}

}

\subsubsection{Integer Programming Formulation}
Because of the deployment constraints, the optimization problem is now combinatorial in nature where all possible deployments make up the discrete set of feasible solutions. A rigorous mathematical formulation now follows. A grid has $W$ candidate grid points, with each point $w = 1, \dots, W$ involving either one or two sub-arrays. Let $N_w \in \{1, 2\}$ denote the number of placements for grid point $w$ and further denote the corresponding placements as $\mathit{\Delta}_w$ and rotations as $\mathcal{R}_w$. Let $x_w \in \{0, 1\}$ denote a selection variable, then the optimization problem can be formulated \begin{equation} \label{eq:OptProblem} \begin{split}
    \min_{x_1, x_2, ..., x_W} & \hspace{6 pt} \rho(\mathcal{A}) \\
    \text{s.t.} & \hspace{3pt} \mathcal{A}_w = \begin{cases}
        \{\mathit{\Delta}_w, \mathcal{R}_w, \mathcal{Q}\} \,, & x_w = 1 \\
        \{\} \,, & \text{otherwise} \end{cases} \\
    & \hspace{3 pt} \mathcal{A} = {\textstyle \bigcup}_w \mathcal{A}_w \\[3pt]
    & \hspace{3 pt} K = \sideset{}{_w} {\textstyle \sum} x_w N_w. \end{split}
\end{equation}

To solve \eqref{eq:OptProblem}, we consider two approaches: (i) an \textit{exhaustive search} over all feasible solutions $\mathcal{A}$; (ii) a low-complexity \textit{greedy search} which starts from an initial deployment of two roof-mounted sub-arrays. In the greedy search, unique sub-arrays are appended to the initial deployment in an iterative manner and according to the placement constraints, where each iteration minimizes $\rho$, until the desired $K$ is met.

\subsection{Understanding the Impact of Ambiguities}
\label{sec:MLObjFun}

The CRLB is a local metric based on the local curvature of the likelihood function around the true value \cite{Kay}. Hence, it cannot capture the behavior outside of this neighborhood of points $\bm{\eta}$. In case of high non-linearity and absence of convexity, the complexity of an estimator employing a $P$-dimensional global search will be significantly increased \cite{Kay}. To understand the impact of specific deployments on the complexity of en estimator and to provide a more complete analysis, we investigate the effects of deployments on the likelihood function in a region around the true value.

Using \eqref{eq:VectorizedObs} and noting that $\yy_1, \dots \yy_K$ are jointly Gaussian and independent, the ML estimator is equivalent to minimizing the sum-of-squares function \begin{equation}
     L(\param) \doteq \sideset{}{_k}\sum \| \yy_k - \sideset{}{_\ell}\sum \vecobsmean{k, \ell} \|^2
 \end{equation} with respect to $\param$ \cite{Kay} and is non-linear and
 generally not convex. Minimizing $L(\param)$ requires numerical methods, and its compressed version is now derived but only for the LOS+GR case. The mean received signal in \eqref{eq:VectorizedObsSignalPart} is rewritten as \begin{equation} \label{eq:MeanExpressedDifferently}
    \sideset{}{_\ell}\sum \vecobsmean{k, \ell} = \CC_k(\pp, \ss) \FF_k(\pp, \ss) \alal_k \,,
\end{equation} 
where $\CC_k(\pp, \ss) = [\cc_{k, 0} \hspace{3pt} \cc_{k, 1}]$ with $\cc_{k, \ell} = \sqrt{P_\mathrm{tx}}(\bb_{k, \ell} \odot \xx) \otimes \aa_{k, \ell} \in \theComplex^{M \hspace{-1.5 pt} N \times 1}$, $\FF_k(\pp, \ss) = \mathrm{diag}( e^{j\phi_{k, 0}}, 1)$ and $\alal_k = \transpose{[\alpha_{k, 0}, \gamma_{k, 1}]}$ where $\gamma_{k, 1} = \alpha_{k, 1}e^{j\phi_{k, 1}} \in \theComplex$. In the coherent mode, $\CC_k(\pp, \ss)\FF_k(\pp, \ss) = \CF_k(\pp, \ss)$ with columns $[\bm{\Omega}_k]_{:,\ell} = \bm{\omega}_{k, \ell}$ and the nuisance parameters can be estimated \begin{equation} \hat{\underline{\alal}}_k(\pp, \ss) = \big( \transpose{\underline{\CF}}_k \underline{\CF}_k \big ) ^{-1} \transpose{\underline{\CF}}_k \underline{\yy}_k \,,
\end{equation} where $\underline{\CF}_k$ is a real matrix given by
\begin{equation}
    \underline{\CF}_k = \begin{bmatrix}
        \Re \bm{\omega}_{k, 0} & \Re \bm{\omega}_{k, 1} & -\Im \bm{\omega}_{k, 1} \\
        \Im \bm{\omega}_{k, 0} & \Im \bm{\omega}_{k, 1} & \Re \bm{\omega}_{k, 1}
    \end{bmatrix} \in \theReals^{2M \hspace{-1 pt} N \times 3} \,,
\end{equation} 
$\underline{\yy}_k = \transpose{[\Re \transpose{\yy_k}, \Im \transpose{\yy_k}]}$ and $\hat{\underline{\alal}}_k = \transpose{[\alpha_{k, 0}, \Re \gamma_{k, 1}, \Im \gamma_{k, 1}]}$. The compressed objective function becomes\begin{equation} \label{eq:LoglikeliCoh}
    L_\mathrm{coh}(\pp, \ss) = \sideset{}{_k} \sum \| \yy_k - \CF_k(\pp, \ss) \bm{\Pi} \hat{\underline{\alal}}_k(\pp, \ss) \|^2,
\end{equation} now independent of $\alal$ and where $\bm{\Pi} \in \theComplex^{2 \times 3}$ transforms $\hat{\underline{\alal}}_k$ into $\hat{\alal}_k$. In incoherent mode, $\FF_k(\pp, \ss) \alal_k = \gamgam_k \in \theComplex^{2 \times 1}$ are arbitrary complex channel coefficients and \begin{equation} \label{eq:LoglikeliIncoh}
    L_\mathrm{incoh}(\pp, \ss) = \sideset{}{_k} \sum \| \yy_k - \CC_k(\pp, \ss) \hat{\gamgam}_k(\pp, \ss) \|^2
\end{equation} for $\hat{\gamgam}_k(\pp, \ss) = (\CC_k^H\CC_k)^{-1}\CC_k^H \yy_k$.

\section{Results \& Discussion}

\subsection{Simulation Parameters}

The channel gains $\alpha_{k, \ell}$ are generated as
\begin{align}
    & \alpha_{k, 0} = \sqrt{G_\mathrm{tx} G_\mathrm{rx, 0}} \frac{\lambda}{4\pi d_{k, 0}} \\
    & \alpha_{k, 1} = |\Gamma_{k, 1}| \sqrt{G_\mathrm{tx} G_\mathrm{rx, 1}} \frac{\lambda}{4\pi d_{k, 1}} \label{eq:GRGain}
\end{align} where $G_\mathrm{tx}$ is a constant antenna gain at the BS and $G_\mathrm{rx, \ell} = G_\mathrm{rx}(\taz_{k, \ell}, \tel_{k, \ell})$ is the antenna gain of each sub-array element. The latter quantity is modeled as \begin{equation}
    G_\mathrm{rx}(\taz_{k, \ell}, \tel_{k, \ell}) = G_\mathrm{max}\cos^{2\beta} \taz_{k, \ell} \cos^{2\beta} \tel_{k, \ell} + G_\mathrm{min} \,,
\end{equation} for $|\taz_{k, \ell}|, \hspace{1 pt} |\tel_{k, \ell}| \leq \ang{90}$ and otherwise zero. The directivity is $G_\mathrm{max} = 10^{8/10}$ (\SI{8}{\decibel i}), $G_\mathrm{min} = 10^{-4}$ (\SI{-40}{\decibel i}) is the minimum gain and $\beta \approx 2.03 $ such that the half-power beam width is $\ang{65}$, and is consistent with the 3GPP model of a patch-antenna element at FR2 in \cite{AntPattern} apart from a smoother taper. The reflection coefficient $\Gamma_{k, 1}$ is calculated using the Fresnel equations \cite[(8-221)]{DKCheng}. The ground is modeled using a complex relative permittivity of $\epsilon = 5.0 + 0.2j$.

Vehicle self-occlusions are detected and implemented through two mechanisms. Initially, if either $|\taz_{k, \ell}|, |\tel_{k, \ell}| > \SI{90}{\degree}$ then the virtual source is outside the sub-array's field-of-view and occluded. Additionally, if there is an intersection along the line spanned by $\tilde{\bm{u}}_{k, \ell}$ with the vehicle body (shown in Fig.~\ref{fig:Grid}), the corresponding path is occluded.

To compute $\rho(\mathcal{A})$, the parameters $(\varphi, r)$ are restricted by considering an example traffic scenario illustrated in Fig.~\ref{fig:RotIllustration}. We seek to guarantee positioning performance indicated by $\rho(\mathcal{A})$ within a certain radius $r_\mathrm{max} = \SI{77}{\meter}$ and outside $r_\mathrm{min} = \SI{5}{\meter}$. $\mathrm{PEB}(\varphi, r; \mathcal{A})$ is sampled at vehicle positions $\pp = \transpose{[0, r, -h_\mathrm{BS} + 0.2]}$ for $r = r_\mathrm{min}, r_\mathrm{min} + \Delta r, \dots, r_\mathrm{min} + (N_r - 1)\Delta r$ with $N_r = 20$ and $\Delta r = \SI{3.6}{\meter}$. Because of placement constraint ii) in Section \ref{sec:OptMethod}, $\mathrm{PEB}(\varphi, r; \mathcal{A})$ is symmetric with respect to $\varphi$ and samples are only generated for a $\ang{180}$ interval $\varphi \in [-\ang{90}, \ang{90})$ with steps of $\Delta \varphi = \ang{3}$ such that $N_\varphi = 60$. The total number of samples are $N_rN_\varphi = 1200$. The remaining simulation parameters are presented in Table \ref{tab:SimParam}.

{

\setlength{\abovecaptionskip}{11 pt}

\begin{table}
    \caption{Simulation parameters.}
    \label{tab:SimParam}
    \centering
    \begin{tabular}{llr}
         \textbf{Parameter} & \textbf{Description} & \textbf{Value}  \\ \hline \hline

         $h_\mathrm{BS}$ & BS height & \SI{20.0}{\meter} \\
         $P_\mathrm{tx}$ & BS average transmit power & \SI{1}{\watt} \\
         $G_\mathrm{tx}$ & BS antenna gain & \SI{10}{\decibel i} \\
         $G_\mathrm{max}$ & Sub-array element directivity & \SI{8}{\decibel i} \\
         $f_c$ & Carrier frequency & \SI{28}{\giga \hertz} \\
         $\Delta_f$ & Subcarrier spacing & \SI{120}{\kilo \hertz} \\
         $N$ & \# subcarriers & \SI{792}{} \\
         $N\Delta_f$ & Bandwidth & \SI{95.0}{\mega \hertz} \\
         $\sigma^2$ & Noise variance & \SI{3.81e-12}{\watt} \\
         $\varepsilon$ & PEB percentile & 0.1 \\
         \hline
    \end{tabular}
\end{table}

}

\subsection{Optimization Results}

The exhaustive search is initiated by considering all possible deployments for the grid, of which there are $\sum_{i = 1}^{20} {20 \choose i} = 1 048 575$ in total, and selecting only the deployments that satisfy the constraint on $K$ to be further processed. For each selected deployment, the metric $\rho(\mathcal{A})$ is calculated. If the FIM $\II_{\param}$ is singular or severely ill-conditioned for any $(\varphi, r)$, the corresponding deployment is deemed to not be identifiable and is discarded. Six separate optimization trials were performed: (i) the \textit{exhaustive} search was run with the different grids mentioned in Section \ref{sec:OptMethod} such that $M=4$ or $M=8$ is varied for both modes; (ii) as a low-complexity alternative, the \textit{greedy} search was performed for the $M=4$ grid, also for both modes. In each trial, the total number of elements $K\hspace{-1 pt}M = 48$ is fixed.

The produced deployments are visualized in Fig.~\ref{fig:FullModelCohResults} and Fig.~\ref{fig:FullModelIncohResults} for coherent and incoherent modes, respectively, and their empirical complementary cumulative distribution functions (ECCDFs) are plotted. In the plots, dashed lines are the results from simulations with the ground reflections turned off. For the exhaustive algorithm and $K=12, M=4$, a total of 48410 and 48339 identifiable deployments were examined of 48412 deployments in total for coherent and incoherent mode, respectively. For $K=6,M=8$, the exhaustive search examined 1898 identifiable deployments out of 1940. The greedy search examined only 80 identifiable deployments. 

In general, the coherent mode deployments promises sub-millimeter positioning error, which is considerably better than the centimeter-level error for incoherent mode. Judging by the dashed lines from the LOS-only model, the ground reflections are detrimental for positioning performance. However, for incoherent mode, the deployments that apply $M=8$ elements seem to fare better compared to $M=4$ in this aspect which is attributed to the improved angular resolution for individual sub-arrays aiding in resolving the paths. The greedy algorithm yields suboptimal deployments as expected and the results are included as a baseline for the optimal exhaustive-search. Interestingly, for coherent mode, the most performant deployment has $K=12, M=4$ and the sub-arrays are visibly spread out to maximize the total near-field coherent aperture. The converse case is seen for incoherent mode results; now the $K=6, M=8$ deployment achieves the smallest PEBs\footnote{In the incoherent mode, the aperture size of the individual sub-arrays plays a more critical role in determining localization accuracy due to the absence of phase synchronization. Larger aperture sizes yields finer angular resolution which, in turn, yields more accurate angle estimates.}. For the incoherent-mode deployments in Fig.~\ref{fig:FullModelIncohResults}, we note the phenomenon that, for $M = 4$, sub-arrays concentrate on the roof. It is thought that the metric $\rho$ is less sensitive to the size of the total array in these cases, and deployments rather seek to maximize received power.

{
\setlength{\abovecaptionskip}{0 pt}

 \begin{figure}
 \centering
  \begin{subfigure}[t]{0.99\columnwidth}
    \centering
    \includegraphics[width= \textwidth, trim= 10 10 10 10, clip]{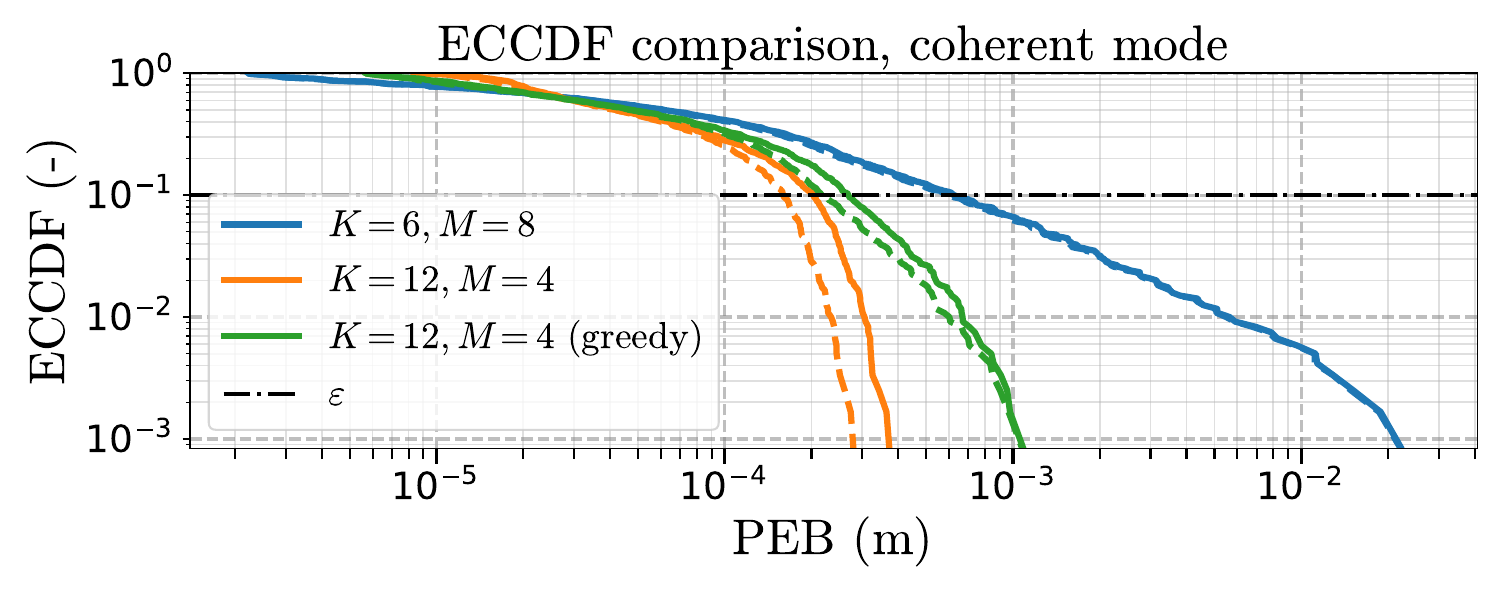}
 \end{subfigure}
 \hfill
 \begin{subfigure}[t]{0.9\columnwidth}
    \centering
    \includegraphics[width= \textwidth, trim= 0 10 0 40, clip]{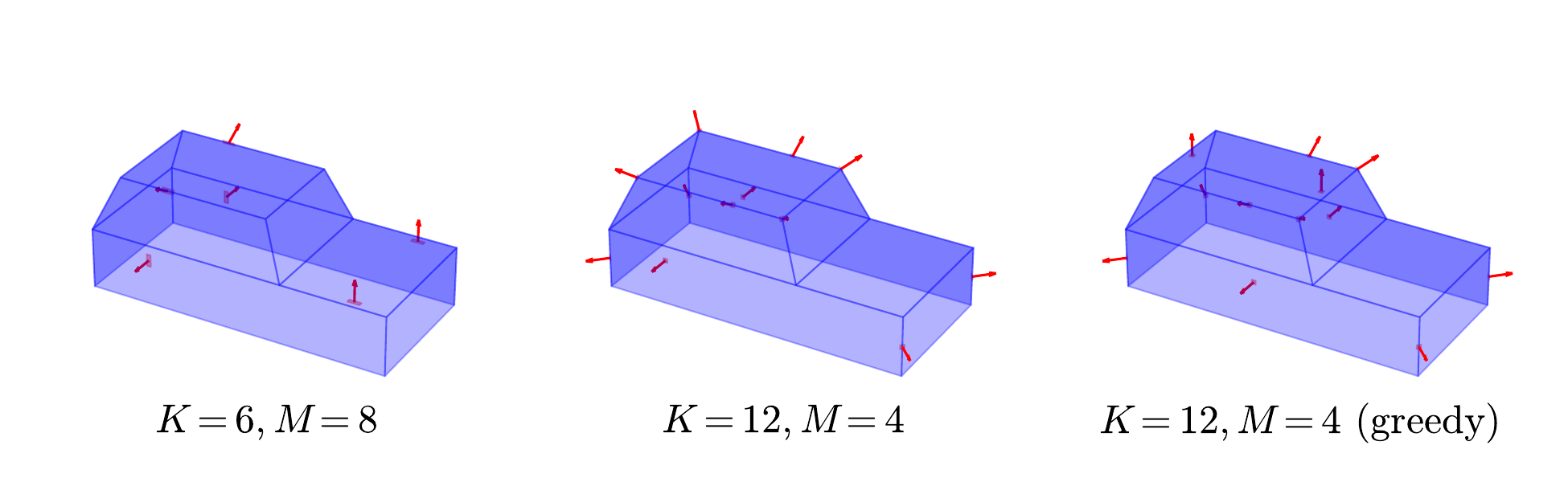}
 \end{subfigure}
 \hfill
 \caption{Optimal deployments and corresponding ECCDFs of $\mathrm{PEB}(\varphi, r;\mathcal{A})$ for coherent mode with $K \hspace{-1 pt} M = 48$. Dashed lines indicate LOS-only model.}
\label{fig:FullModelCohResults}
\end{figure}

 \begin{figure}
 \centering
  \begin{subfigure}[t]{0.99\columnwidth}
    \centering
    \includegraphics[width= \textwidth, trim= 10 10 10 10, clip]{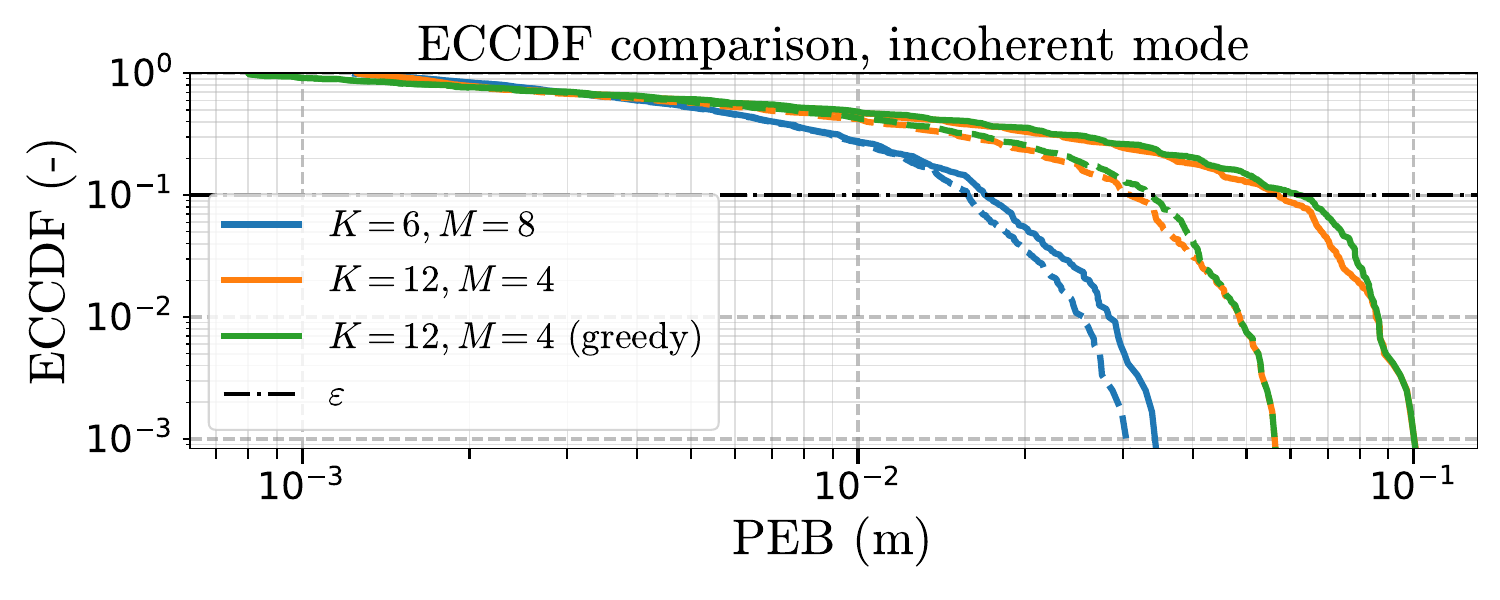}
 \end{subfigure}
 \begin{subfigure}[t]{0.9\columnwidth}
    \centering
    \includegraphics[width= \textwidth, trim= 0 10 0 40, clip]{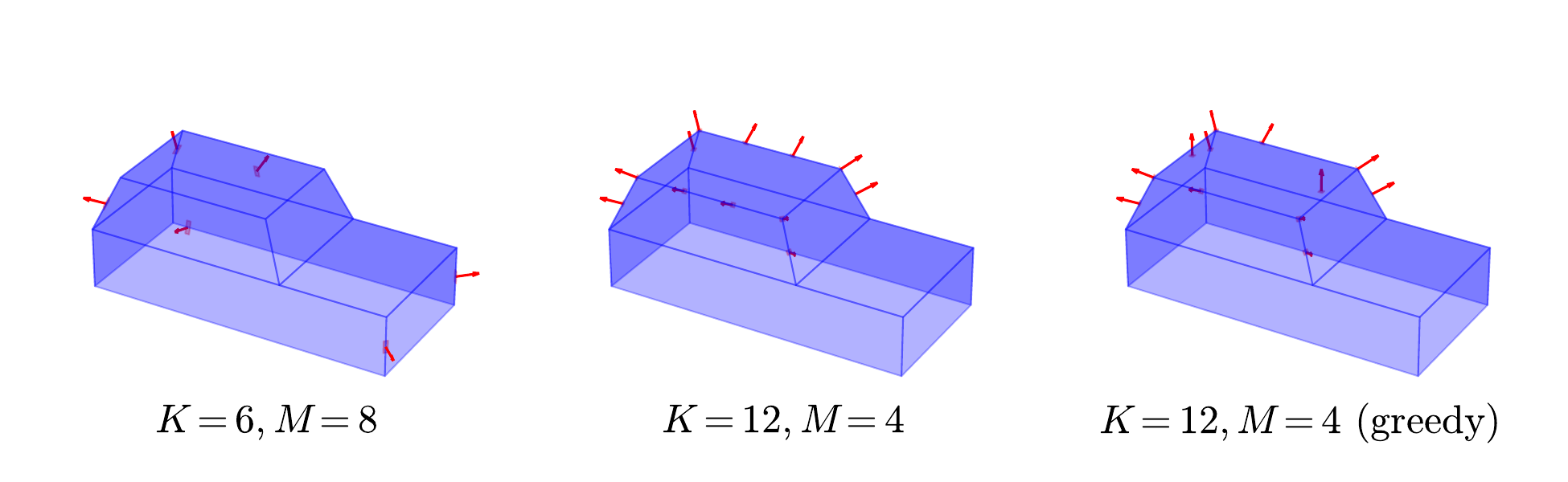}
 \end{subfigure}
 \hfill
 \caption{Optimal deployments and corresponding ECCDFs of $\mathrm{PEB}(\varphi, r;\mathcal{A})$ for incoherent mode with $K \hspace{-1 pt} M = 48$. Dashed lines indicate LOS-only model.}
\label{fig:FullModelIncohResults}
\end{figure}

}

\subsection{Implications for an ML Estimator}

Next, the ML objective functions generated from \eqref{eq:LoglikeliCoh} and \eqref{eq:LoglikeliIncoh} are drawn in the top and bottom of Fig.~\ref{fig:Loglikeli} for coherent and incoherent modes, respectively, using optimal deployments $K = 12, M=4$ for coherent mode and $K = 6, M=8$ for incoherent mode. Now the vehicle $y$-coordinate is set to $r=\SI{25}{\meter}$ and $\varphi= \ang{30}$. The figures depict the noise-less case and $\sqrt{\cdot}$ is applied for visualization purposes. Evidently, $L_\mathrm{incoh}(x, y)$ has a visibly clear minima centered at the true position and is locally smooth. However, for $L_\mathrm{coh}(x, y)$ the minima is not clearly visible due to a small-scale ripple causing aliasing effects. Functions are sampled at $1\lambda$ intervals.

Cuts in the $x$- and $y$ direction at respective true coordinates are shown in Fig.~\ref{fig:Cuts}, now sampled with a finer interval of $\lambda/10$. The zoomed-in portion of the $y$-cut (right) in Fig.~\ref{fig:Cuts} use $\lambda/100$ sample spacing. 

Examining the left $x$-cut reveals that the notably enhanced performance of coherent processing over incoherent processing can be attributed to a sharp global minimum. However, this minimum is poorly positioned among a multitude of local minima. For the $y$-cut, coherent processing can yield estimates that are ambiguous with some value close to $\lambda/2$ ($\approx \SI{5}{\milli \meter}$ in this case), even when the initial guess is near the true position, as shown in the zoomed-in section of Fig.~\ref{fig:Cuts} (right).

{

\setlength{\abovecaptionskip}{-11 pt}
\setlength{\belowcaptionskip}{-11 pt}

 \begin{figure}
 \centering
 \begin{subfigure}[t]{0.99\columnwidth}
    \centering
    \includegraphics[width= \textwidth]{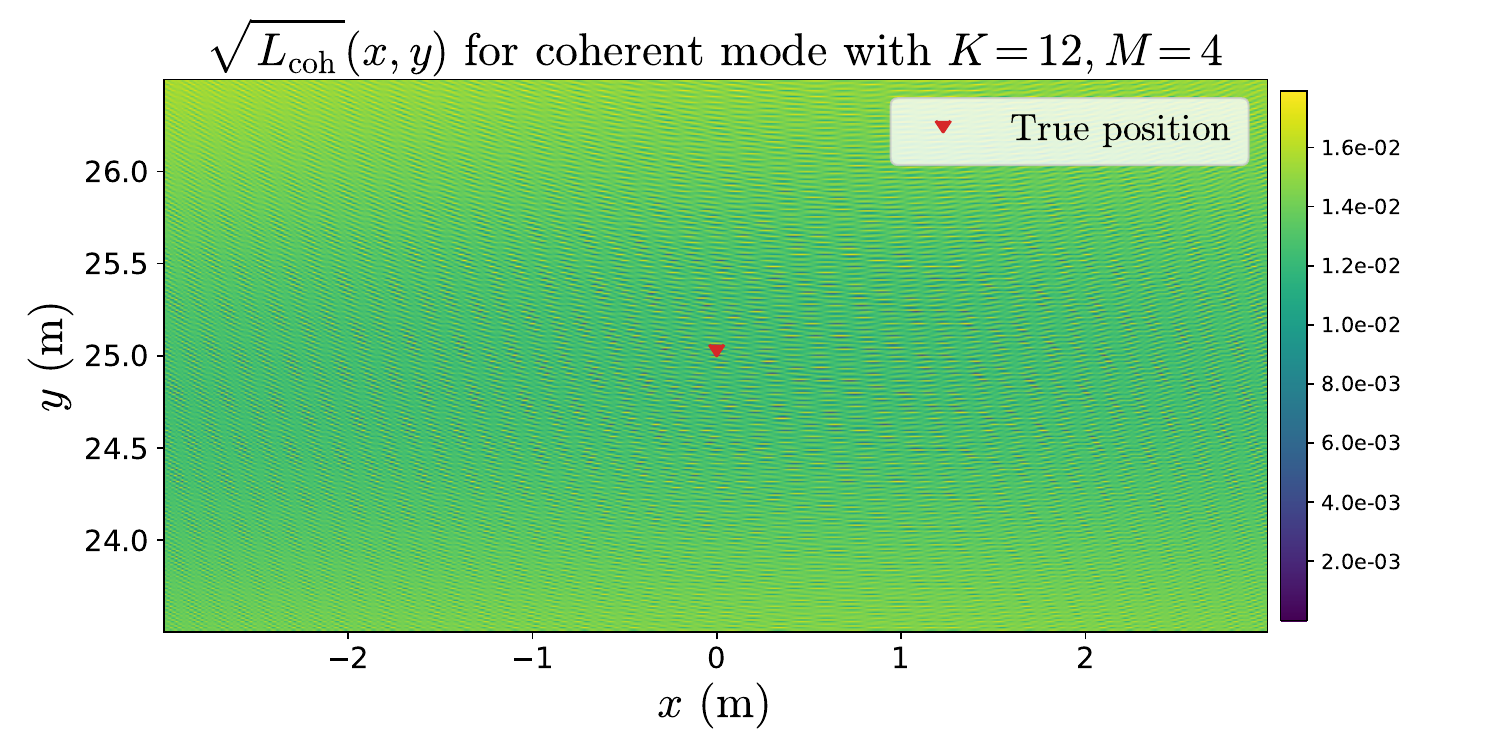}
 \end{subfigure}
 \hfill
 \begin{subfigure}[t]{0.99\columnwidth}
    \centering
    \includegraphics[width= \textwidth]{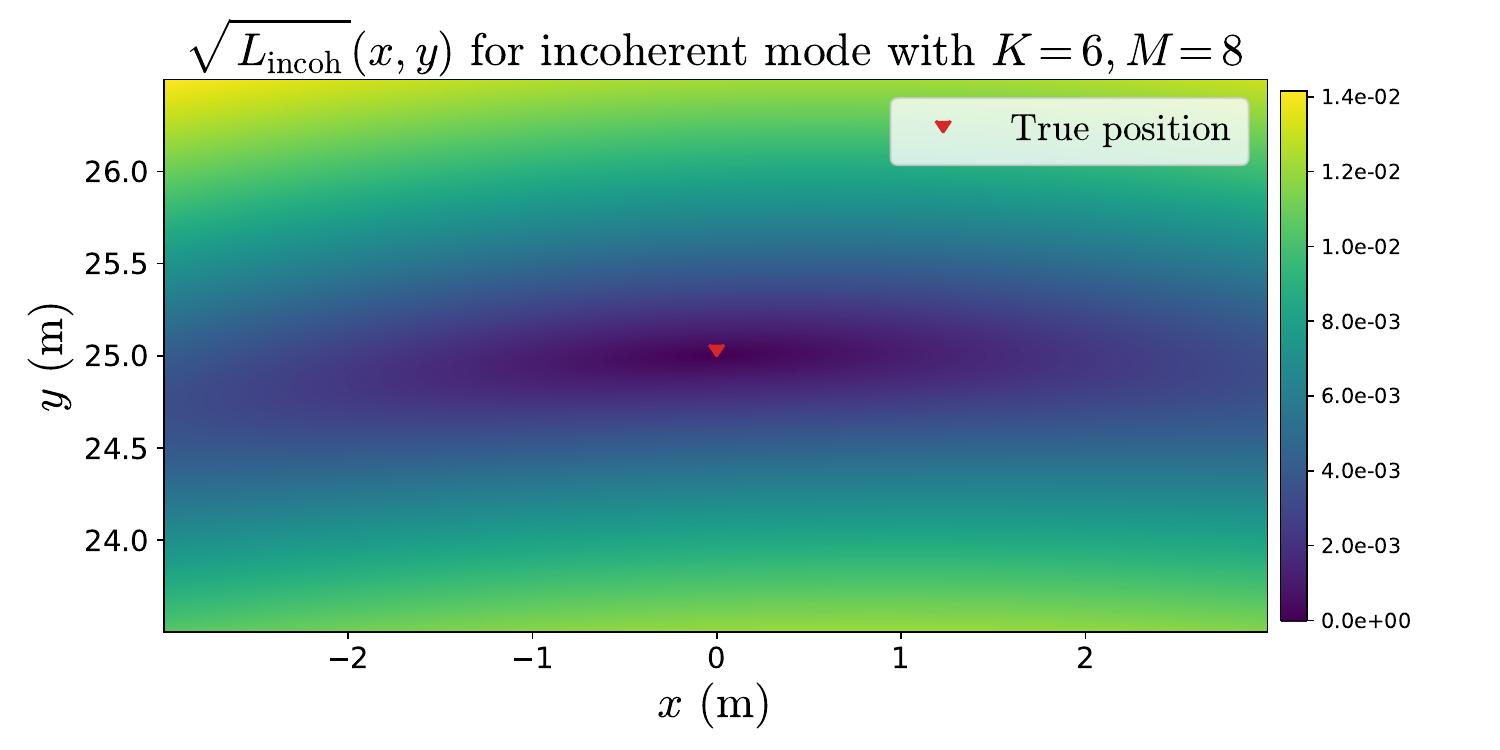}
 \end{subfigure}
 \hfill
 \caption{$\sqrt{L(x, y)}$ for coherent (top) and incoherent (bottom). Both deployments are optimal with $K \hspace{-1pt} M = 48$.}
\label{fig:Loglikeli}
\end{figure}

}
{

\setlength{\abovecaptionskip}{-11 pt}

\begin{figure}
\centering
\begin{subfigure}[t]{0.49\columnwidth}
    \centering
    \includegraphics[width= \textwidth]{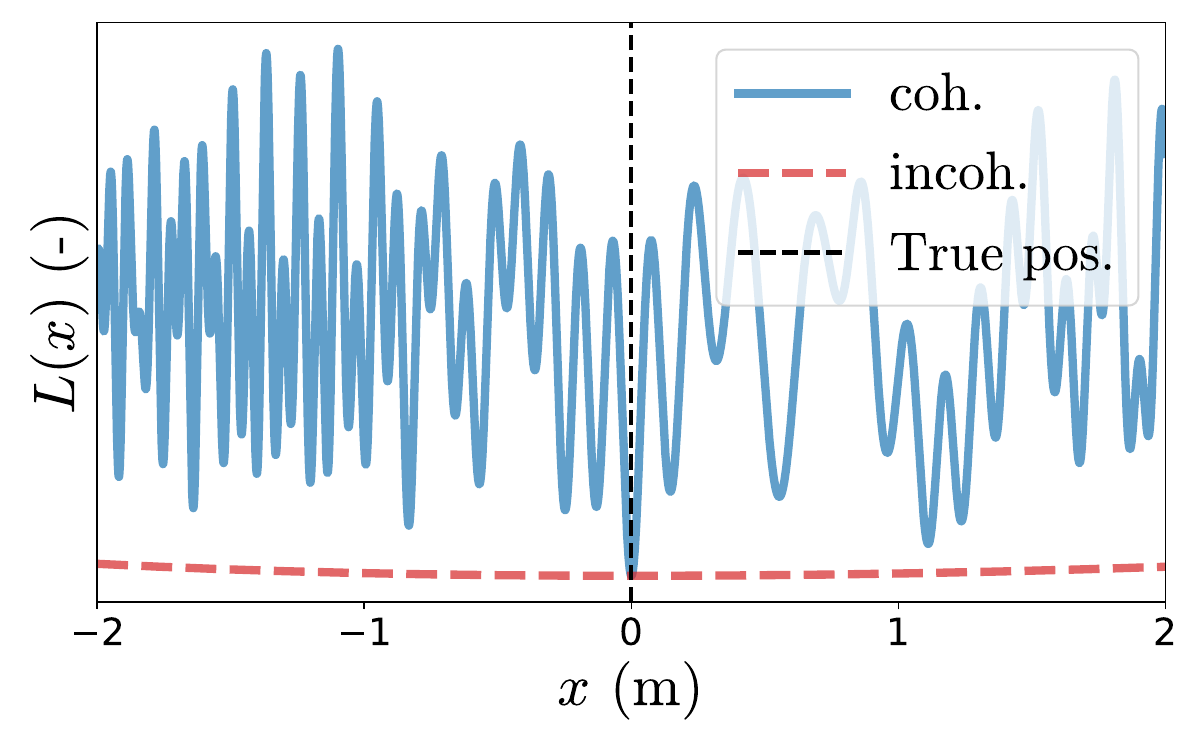}
 \end{subfigure}
 \hfill
  \begin{subfigure}[t]{0.49\columnwidth}
    \centering
    \includegraphics[width= \textwidth]{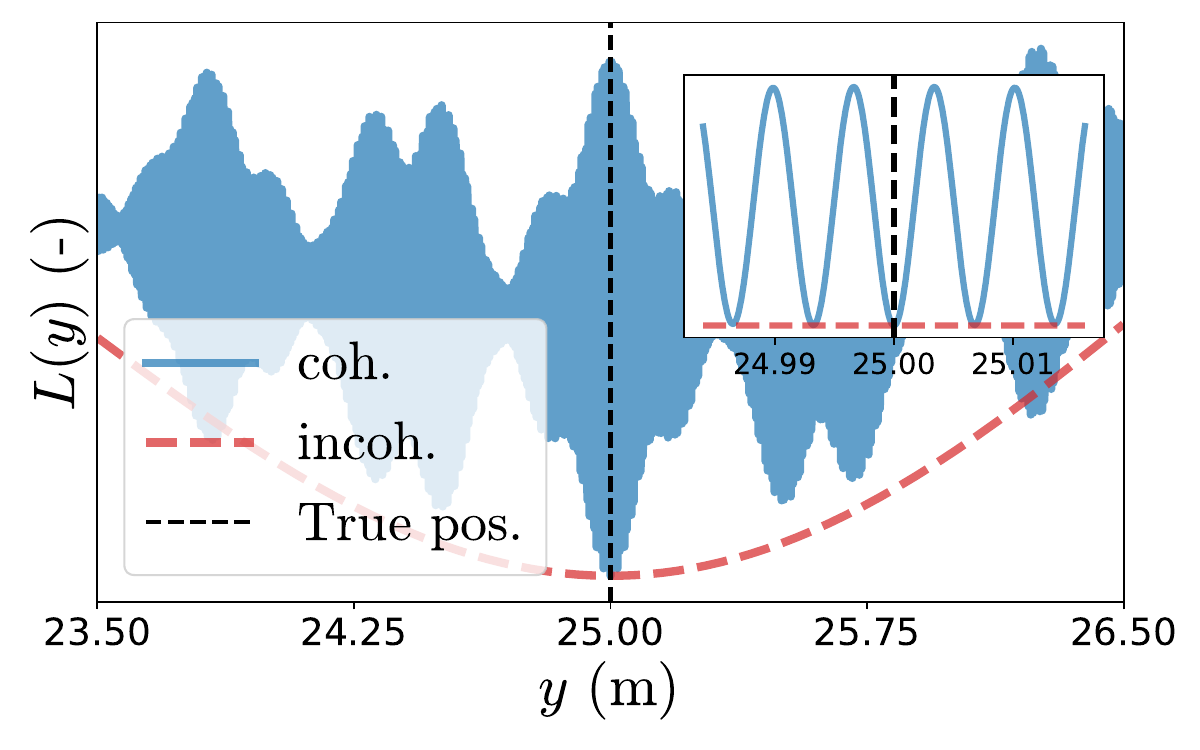}
 \end{subfigure}
 \hfill
 \caption{Individual cuts in $x$ and $y$ of $L(x, y)$ in Fig.~\ref{fig:Loglikeli} but sampled finer.}
\label{fig:Cuts}
\end{figure}

}

\section{Conclusion}

For the positioning-oriented sub-array deployment optimization problem under consideration, an exhaustive search is guided by a set of geometrical constraints to produce optimal placements of sub-arrays whilst taking into account self-occlusion effects and specular multipath. It shows that a large total aperture is preferred for coherent processing with phase synchronization. If the arrays are not phase synchronized, the individual array size and, in turn, the individual angular resolution are more important. Coherent processing delivers sub-millimeter positioning accuracy for the simulated FR2-based scenario, outperforming the centimeter-level accuracy of incoherent processing, but requires a more complex ML estimator to achieve this performance.

\bibliographystyle{IEEEtran}
\bibliography{IEEEabrv,references}

\end{document}